\documentclass[aps,prl,twocolumn,amsmath,amssymb,floats,floatfix,showpacs]{revtex4}

\usepackage{graphicx}% Include figure files
\usepackage{dcolumn}% Align table columns on decimal point

\begin{document}

\title{Non-Markovian quantum jumps} 

\author{Jyrki Piilo}
%\email{jyrki.piilo@utu.fi}
\affiliation{
Department of Physics, University of Turku, 
FI-20014 Turun yliopisto, Finland
}

\author{Sabrina Maniscalco}
\affiliation{
Department of Physics, University of Turku, 
FI-20014 Turun yliopisto, Finland
}

\author{Kari H\"ark\"onen}
\affiliation{
Department of Physics, University of Turku, 
FI-20014 Turun yliopisto, Finland
}

\author{Kalle-Antti Suominen}
\affiliation{
Department of Physics, University of Turku, 
FI-20014 Turun yliopisto, Finland
}

\date{\today}

\begin{abstract}
Open quantum systems that interact with structured reservoirs exhibit non-Markovian dynamics. We present a quantum jump method for treating the dynamics of such systems. This approach is a generalization of the standard Monte Carlo Wave Function (MCWF) method for Markovian dynamics. The MCWF method identifies decay rates with jump probabilities and fails for non-Markovian systems where the time-dependent rates become temporarily negative. Our non-Markovian quantum jump (NMQJ) approach circumvents this problem and provides an efficient unravelling of the ensemble dynamics.

\end{abstract}


\maketitle

{\it Introduction.}
When an open quantum system interacts with a reservoir having non-trivial structure, the system dynamics exhibits non-Markovian memory effects~\cite{Breuer2002}. The information on the state of the open system is contained in the density matrix whose time evolution is governed by a master equation consisting of two parts. The system Hamiltonian induces unitary evolution of the density matrix while the dissipative part, which includes the information on the properties of the environment in form of decay rates, induces non-unitary effects via the jump operators. Already for Markovian systems, which do not have memory, finding the solution of the master equation may be very complicated. The task gets even more challenging with non-Markovian systems and structured reservoirs.
Such systems display modified decay dynamics paving the way
to new types of quantum control techniques~\cite{Lambro}.
 
Non-Markovian systems appear in many branches of physics, such as quantum optics~\cite{Breuer2002,Lambro,Gardiner96a}, solid state physics~\cite{SS}, and quantum chemistry~\cite{QC}. In quantum information processing~\cite{Stenholm2005}, the non-Markovian character of decoherence has to be accounted
for and it leads to the concept of non-Markovian quantum channels~\cite{QIP}. Decoherence also plays a central role in the transition from quantum to classical 
world~\cite{Zurek}. In fact, non-Markovianity has been recently proposed as a means to manipulate the quantum-classical border~\cite{Maniscalco2006}. 
Since it is elusive to solve the open system dynamics, new methods for non-Markovian systems are highly desired.
 
In this Letter we provide an efficient way to unravel a general  non-Markovian master equation. The different ways to build an ensemble of stochastic wave functions describing the density matrix fall roughly into two categories~\cite{Carmichael1993}: time-evolution including (i) discontinuous changes (quantum jumps), e.g., the Monte Carlo Wave Function (MCWF) method~\cite{DCM1992}; (ii) continuous stochastic changes, e.g.,~the Quantum State Diffusion (QSD) method~\cite{Strunz1999,Percival}. Our non-Markovian quantum jump (NMQJ) method generalizes the widely used Markovian MCWF into the field of non-Markovian systems, and thus belongs to the first of  the two categories.

There exists a non-Markovian variant  of QSD~\cite{Strunz1999} and a somewhat related formulation~\cite{Grabert02}.
These methods, however, 
are difficult to implement beyond very simple examples.
Other unravelings of non-Markovian master equations contain 
fictitious harmonic oscillator modes~\cite{Imamoglu} and pseudomodes~\cite{Garraway1996}, or some other forms of extensions of the
system Hilbert space~\cite{BreuerGen, Breuer99}. 
One formulation, using quantum jumps, exploits an analogue to the the hidden variable theory~\cite{Gambetta2003}.  The use of extended Hilbert spaces comes always with an added cost for computational efficiency. 
 
Our formulation avoids the use of Hilbert space extensions and is based on the following observation.  The information, which the system loses to the environment at the time of the jump, can be later recovered by the system due to non-Markovian memory. We show explicitly how this happens on the level of single realizations. Before discussing on the insight and benefits that our NMQJ method provides, we first introduce the master equation and the method, and present a case study with an atom in a photonic band gap.

{\it Non-Markovian master equation.} The non-Markovian dynamics of the reduced system density matrix $\rho(t)$ is given by the master equation~\cite{Breuer2002}
\begin{eqnarray}
   \dot{\rho}(t) &=&  \frac{1}{i\hbar} \left[ H_s, \rho(t)\right] + 
   \sum_j\Delta_j(t) C_j(t) \rho(t)C_j^{\dag}(t) \nonumber \\
  & -&\frac{1}{2}\sum_j\Delta_j(t)\left\{\rho(t),C_j^{\dag}(t) C_j(t) \right\}.
    \label{Eq:MNM}
\end{eqnarray}
Above, $H_s$ is the system Hamiltonian and $C_j(t)$ are the jump operators describing changes in the system due to interaction with the reservoir. $\Delta_j(t)$ is the decay rate of channel $j$.  
It can be shown that the most general master equations local in time for non-Markovian systems can be cast in the form of Eq.~(\ref{Eq:MNM})~\cite{BreuerGen}. 
In the Markovian case all $\Delta_j$ are positive constants. In the non-Markovian case the rates may oscillate 
and take negative values for finite time intervals. This is a sign of  the non-Markovian memory effects and reflects the exchange of information back and forth between the system and the reservoir.

{\it MCWF and NMQJ methods.} 
The system properties are calculated as an average over the state vector ensemble of size $N$ and  we follow closely
the MCWF method~\cite{DCM1992}.
A generic way to write the density matrix
 % in terms of the ensemble 
 is 
\begin{equation}
\label{Eq:Rho}
\rho(t) = \sum_\alpha \frac{N_\alpha(t)}{N} |\psi_\alpha(t)\rangle \langle \psi_\alpha(t)|,
\end{equation}
where $N_\alpha(t)$ is the number of ensemble
members in the state $|\psi_\alpha(t)\rangle$ at time $t$.
The deterministic evolution of a given state vector $|\psi_\alpha(t)\rangle$, for small enough time steps $\delta t$ and before the renormalization, is given by
\begin{equation}
\label{Eq:Det}
| \phi_\alpha(t+\delta t)\rangle = 
 \left(1-\frac{iH\delta t}{\hbar}\right)  |\psi_\alpha(t)\rangle,
\end{equation}
where the non-Hermitian Monte Carlo Hamiltonian is $H= H_s-i\hbar\sum_j\frac{1}{2}\Delta_j(t)C_j(t)^{\dagger}C_j(t)$
and the renormalized state is $ |\psi_\alpha(t+\delta t)\rangle = |\phi_\alpha(t+\delta t)\rangle / |||\phi_\alpha(t+\delta t)\rangle ||$.
For positive decay channels $j_+$, $\Delta_{j_+}(t)>0$,  the deterministic evolution is  interrupted by jumps 
$|\psi_\alpha(t)\rangle  \rightarrow  C_{j_+}(t) |\psi_\alpha(t)\rangle  / ||C_{j_+}(t)| \psi_\alpha(t)\rangle||$ which occur 
with probability
\begin{equation}
 P_\alpha^{j_+}(t)=\Delta_{j_+} (t)\delta t \langle \psi_\alpha(t) | C_{j_+}^{\dagger}(t)C_{j_+}(t)|\psi_\alpha(t)\rangle,
\end{equation}
during time step $\delta t$~\cite{DCM1992}.
The Markovian MCWF method can be extended to the situations where the rates become time dependent, but this is limited
to positive decay rates only.

In our approach the non-Markovian quantum jumps for negative channels $j_-$, $\Delta_{j_-}(t)<0$, have the form 
\begin{equation}
D_{\alpha\rightarrow \alpha'}^{j_-}(t)= |\psi_{\alpha'}(t)\rangle \langle \psi_{\alpha}(t)|,
\label{Eq:JOp}
\end{equation}
where the source state of the jump is $|\psi_{\alpha}(t)\rangle = C_{j_-}(t) |\psi_{\alpha'}(t)\rangle / ||C_{j_-}(t) |\psi_{\alpha'}(t)\rangle||$.
This transition for a given state vector $|\psi_\alpha\rangle$ in the ensemble (\ref{Eq:Rho}) occurs with the probability 
\begin{eqnarray}
   P_{\alpha\rightarrow \alpha'}^{j_-}(t) &=& \frac{N_{\alpha'}(t)}  {N_{\alpha}(t)} |\Delta_{j_-}(t)| \delta t  
 \nonumber \\
   &\times&
   \langle \psi_{\alpha'}(t) | C_{j_-}^{\dagger}(t) 
   C_{j_-}(t) |\psi_{\alpha'}(t)\rangle.
   \label{Eq:JProb}
\end{eqnarray}
Note that the probability of the non-Markovian jump is given by the target state of the jump. 
The sign of the decay rate $\Delta_j(t)$ can be understood in the following way. First, when for a given channel $j$,
$\Delta_j(t)>0$,  the process goes as  $|\psi_{\alpha'}\rangle\rightarrow |\psi_{\alpha}\rangle =C_{j}|\psi_{\alpha'}\rangle / ||C_{j} |\psi_{\alpha'}\rangle||$.
Later on, when the decay rate becomes negative,  $\Delta_j(t)<0$,  the direction of this process is reversed and the jump occurs to opposite direction $|\psi_{\alpha'}\rangle\leftarrow |\psi_{\alpha}\rangle$.

The proof of our NMQJ method goes in a very similar way to that of the Markovian 
MCWF method~\cite{DCM1992}. By weighting the deterministic and jump paths over time step $\delta t$ with the appropriate probabilities we should obtain the master
equation (\ref{Eq:MNM}). 
Calculating the average $\overline{\sigma}$ of the evolution  of the ensemble (\ref{Eq:Rho}) over
$\delta t$ gives 
\begin{eqnarray}
\label{Eq:Algo}
&&\overline{\sigma}(t+\delta t) =
\sum_\alpha\frac{N_\alpha(t)}{N} 
\left[
\left(
1-\sum_{j_+}P_\alpha^{j_+}(t) \right.\right.
\nonumber \\
&-&
\left.
\sum_{j_-,\alpha'} P_{\alpha\rightarrow \alpha'}^{j_-}(t)
\right)
\frac{| \phi_\alpha(t+\delta t)\rangle \langle \phi_\alpha(t+\delta t) |}{||| \phi_\alpha(t+\delta t)\rangle ||^2}
\nonumber \\
&+&
\sum_{j_+} P_\alpha^{j_+}(t)
\frac{C_{j_+}(t) |\psi_\alpha(t)\rangle \langle \psi_\alpha(t)| C_{j_+}^{\dagger}(t) }
{||C_{j_+}(t) |\psi_\alpha(t)\rangle||^2}
\nonumber \\
&+&
\left.
\sum_{j_-,\alpha'} P_{\alpha\rightarrow \alpha'}^{j_-}(t) D_{\alpha\rightarrow \alpha'}^{j_-}(t) |\psi_\alpha(t)\rangle \langle \psi_\alpha(t)| D_{\alpha\rightarrow \alpha'}^{j_-\dagger}(t)
\right].\nonumber \\
\end{eqnarray}
Here, the summations $\alpha$ and $\alpha'$ run over the ensemble, the summation over $j_+$ and $j_-$ cover 
the positive and negative
channels, respectively. 
The first term on the r.h.s. for the summation over $\alpha$ is the product of the no-jump probability and the deterministic evolution of the state vector,
the second and third terms describe the positive and negative channel jumps, respectively, with the corresponding probabilities.
By using Eq.~(\ref{Eq:JOp}), the last term can also be written 
as 
$\sum_{j_-,\alpha'}P_{\alpha\rightarrow \alpha'}^{j_-}(t)|\psi_{\alpha'}(t)\rangle \langle \psi_{\alpha'}(t)|$.
 In general, using Eqs.~(\ref{Eq:Rho})-(\ref{Eq:JProb}) in Eq.~(\ref{Eq:Algo}),
and keeping in mind the form of the reversed jump $|\psi_\alpha\rangle  \rightarrow|\psi_{\alpha'}\rangle$ with $|\psi_{\alpha}\rangle = C_{j_-}|\psi_{\alpha'}\rangle  / ||C_{j_-} |\psi_{\alpha'}\rangle||$, gives the master equation (\ref{Eq:MNM}).

{\it Example: Photonic band gap.}
To illustrate the NMQJ method we choose a two-level atom inside a photonic band gap (PBG)~\cite{John94, Lambro}. Fictitious and pseudomode methods~\cite{Imamoglu, Garraway1996} do not work for this system since the typical reservoir distribution function for PBG is not a meromorphic function due to the band edge. Moreover, an earlier attempt to develop a specific  jump approach for this system~\cite{Quang97} has been shown to be correct only in the Born-Markov limit~\cite{Molmer97}. One of the reasons for this is that the method
of Ref.~\cite{Quang97} fails to describe the reabsorption of photons by the
atoms~\cite{Molmer97}.
Our method succeeds in this by using non-Markovian quantum
jumps [c.f.~Eq.~(\ref{Eq:JOp})]. This example also shows that local-in-time master equations can be used to  describe non-Markovian dynamics for strong system-reservoir interactions.

\begin{figure}[tb]
\centering
\includegraphics[scale=1.0]{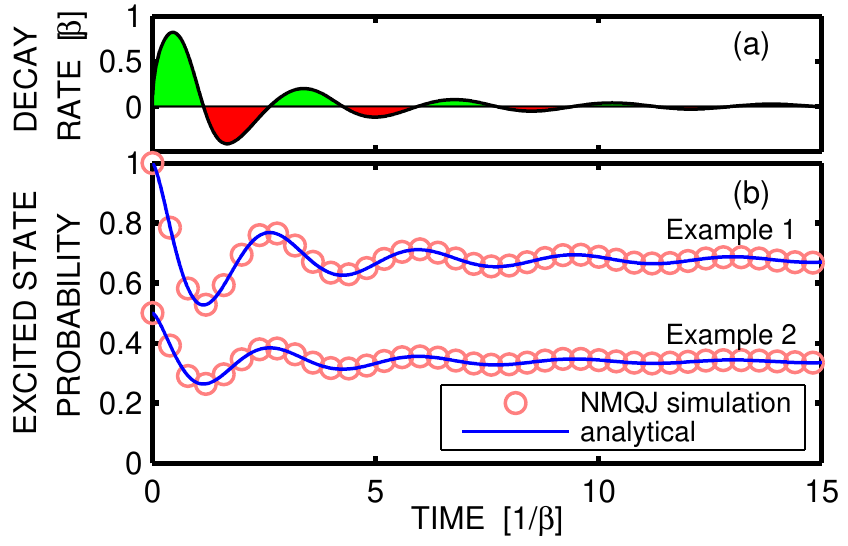}
\caption{\label{Fig:1} 
(Color online) (a) The decay rate for a two-level atom in photonic band gap as a function of time.
(b) NMQJ and exact results. In (a) the decay rate displays oscillatory behavior with temporary negative values. In (b) we plot the excited state probability of the atom and the results show the match between the exact and simulation results. The initial pure states in examples $1$ and $2$ are $|e\rangle$ and $( |g\rangle+ |e\rangle) / \sqrt{2}$, respectively.
}
\end{figure}

The master equation for the density matrix of the two-level system takes the form~\cite{Breuer2002}
\begin{eqnarray}
\label{Eq:ME2}
 \dot{\rho}(t) & = &\frac{1}{i\hbar} \frac{S(t)}{2} \big[\sigma_+ \sigma_-, \rho(t) \big] 
 \nonumber \\
&+&
\Delta(t) \Big( \sigma_- \rho(t) \sigma_+ - \frac{1}{2} \left\{\sigma_+ \sigma_-, \rho(t)\right\} \Big),
\end{eqnarray}
where $S(t)$ is the Lamb shift, $\Delta(t)$ the decay rate, $\sigma_-=|g\rangle\langle e |$, and 
$\sigma_+=\sigma_-^{\dagger}$. Here,  $|g\rangle$ denotes the ground state of the two-level atom,
$|e\rangle$ the excited state, and there is one decay channel taking the atom from $|e\rangle$ to $|g\rangle$.
We calculate the Lamb shift and the decay rates by using Eq.~(2.21) of Ref.~\cite{John94} and 
Eqs.~(10.22) and (10.23) from Ref.~\cite{Breuer2002}. The oscillatory behavior and negative values of the decay rate are displayed in Fig.~\ref{Fig:1} (a).

%%%%%%%%%%%% FIG
%%%%%%%%%%%% FIG

\begin{figure}[tb]
\centering
\includegraphics[scale=1.0]{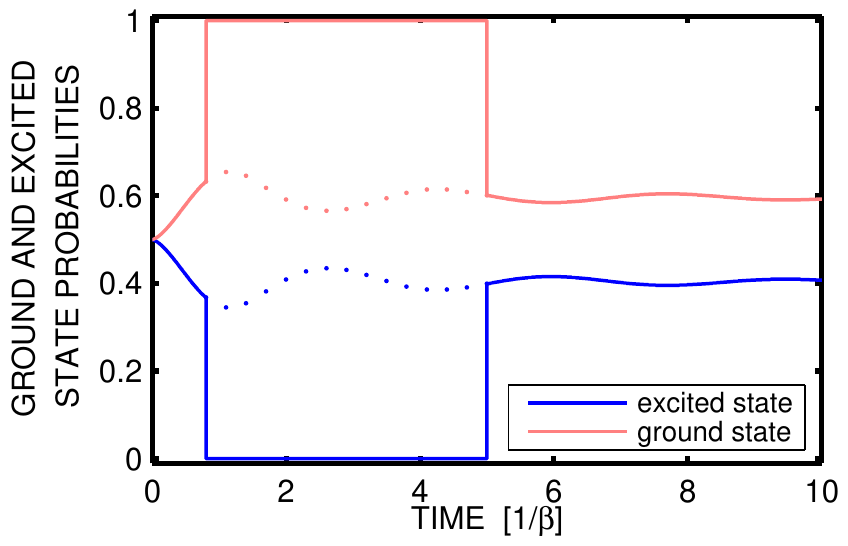}
\caption{\label{Fig:2} 
(Color online)
An example realization with a jump - reverse jump cycle. The ground and excited state probabilities are given as function of time. The first jump at time $t\simeq0.8$ occurs at the positive decay rate region and destroys the superposition state. The second jump at $t\simeq5.0$ occurs at the negative decay region and recreates the superposition. The dotted lines show the evolution without any jumps. 
}
\end{figure}

Figure~\ref{Fig:1} (b) shows the match between the exact result [c.f. Eq.~(2.21) of Ref.~\cite{John94}]
and the simulation results with $N=10^5$ realizations for two different initial states.
We have chosen parameters which correspond to Fig.~1 of Ref.~\cite{John94} with the detuning
$\delta = -\beta$ from the edge of the gap. Here, $\beta=(\omega_{0}^{7/2}d^2 / 6\pi \epsilon_0\hbar c^3)^{2/3}$, where $\omega_0$ is the Bohr frequency and $d$ the absolute value of dipole moment of the atom. The results illustrate a typical feature of PBG: atom-photon bound state and population trapping. 
Figure~\ref{Fig:2} displays an example of non-Markovian quantum jump in a single realization of the process for the case of initial superposition state. First, during the positive decay, a jump takes the atom to its ground state and the excitation resides in the environment. Later on with negative rate, the superposition state is restored by a non-Markovian quantum jump, and the photonic component is reabsorbed by the atom.  

{\it Insight by NMQJ.}
In the PBG example above, the key ingredient to describe non-Markovian
memory is the virtual photon emission-reabsorption cycle on the level
of single realization.
The physical state of the system is given by the density matrix,
i.e., the ensemble of state vectors. 
This illustrates an interesting aspect of our method: it is possible to describe 
the effects of non-Markovian memory without extending the Hilbert space of the reduced system, which is a trait used in the previously developed jump
methods~\cite{Imamoglu, Garraway1996,Breuer99,BreuerGen}.
In NMQJ method, 
the memory of the ensemble member $|\psi_\alpha\rangle$, i.e.,~the information about the state before the positive rate jump to the state $|\psi_\alpha\rangle$ occurred, is carried by the other ensemble member $|\psi_{\alpha'}\rangle$.
Consequently, the density matrix and the corresponding ensemble indeed carry information on the earlier state of the system.

%%%%%%%%%%%% FIG
%%%%%%%%%%%% FIG

Negative decay rates, which occur in non-Markovian
systems, can be interpreted in the following way.
During the initial period of positive decay, the corresponding jumps distribute the state vector probability over the Hilbert space accordingly; the number of terms in the summation of Eq.~(\ref{Eq:Rho}) increases. 
When the decay rate later on becomes negative, which indicates the memory effects, the direction of the probability flow is reversed. 
This means that a process $|\psi\rangle \rightarrow |\psi'\rangle$ with negative rate corresponds to  $|\psi\rangle \leftarrow |\psi'\rangle$.
From classical perspective, it seems rather usual that changing the sign of the rate of the process means that the process occurs to the opposite direction. 
In the quantum world with superpositions, probability amplitudes and coherences the issue is less straightforward.
In our method, this appears as a restoration of seemingly lost superpositions and subsequent revival of coherences.

{\it The algorithm and numerical efficiency.}
Since in the NMQJ method the realizations depend on each other due to memory effects [c.f.~Eq.~(\ref{Eq:JProb})], it seems at first sight that all the $N$ ensemble members have to be evolved simultaneously.  However, according to Eq.~(\ref{Eq:Rho}), the ensemble consists of several copies of each $|\psi_\alpha(t)\rangle$. Obviously, there is no need to have on a computer several copies of the same state vector. It is sufficient to have one copy and the corresponding integer 
number $N_\alpha$.  Any number $N$ of the realizations of the process 
can be done by making $N_{\rm eff}\ll N$ state vector evolutions
where $N_{\rm eff}$ is equal to the number of terms in the summation $N=\sum_\alpha N_\alpha$.
When the realizations of the process are generated on a computer, a jump means changing the integer numbers $N_\alpha(t)$ accordingly in Eq.~(\ref{Eq:Rho}).
A considerable saving in CPU time is achieved since it is not necessary at each point of time evolve $N$ state vectors, instead, it is enough to decide $N$ times if the jumps occurred or not. 

Let us illustrate this with an example.
In the PBG case above, we have $N_{\rm eff}=2$  and the corresponding state vectors are:
$|\psi_0(t)\rangle$ and $|\psi_1(t)\rangle = |g\rangle$ for all $t$. These are the initial state affected by the deterministic evolution and the ground state, respectively. 
In the positive decay region the jumps occur as $|\psi_0(t)\rangle \rightarrow |\psi_1(t)\rangle=|g\rangle$; each jump reduces $N_0$ by $1$ and increases $N_1$ by $1$.
In the negative decay region the process goes to opposite direction $ |\psi_1(t)\rangle=|g\rangle \rightarrow |\psi_0(t)\rangle$.  In the
optimized  simulation to have $10^5$ realizations we need to generate
only one deterministic
evolution for $|\psi_0(t)\rangle$, and  then decide the jumps
as described above. 

In QSD~\cite{Strunz1999},
the stochastic change of the state vectors is continuous which leads
in practice to $N_{\rm eff}=N$. For the doubled Hilbert space (DHS) 
method~\cite{Breuer99}, the norm of the state vectors increase in the negative decay 
region. As a consequence,
the norm of a given state vector depends on the point of time where the DHS jump happens
during the negative decay. In the ensemble, the jumps occur at each time point
and $N_{\rm eff}$ becomes large compared to the NMQJ method.
Moreover, the DHS state vectors are evolved in the Hilbert space twice as large as in NMQJ. 

In contrast to the DHS method, the triple Hilbert space method (THS) preserves the norm of state vectors~\cite{BreuerGen}.
However, in the most general case, when the jump operators depend on time
in the master equation (\ref{Eq:MNM}), the jumps with the extended THS
operators increase $N_{\rm eff}$ at each point of time during the negative decay. 
Consequently, the THS method can not use 
the built-in optimization of the NMQJ method.
Moreover, the THS method has two other ingredients which have an impact on its numerical performance: (i) the need for 4 times larger number of decay channels than NMQJ uses
[see the text below Eq.~(57) in
Ref.~\cite{BreuerGen}], (ii) the state vectors live in the space which is $3$
times larger than the original one [see Eqs.~(27)-(29) in Ref.~\cite{BreuerGen}]. 
The consequent complications of the THS method make it difficult to make a general statement on its numerical performance. 
However, all the facts above lead to the conclusion that even the most cautious estimate would give roughly an order of magnitude difference in the numerical efficiency between the NMQJ and the THS methods. 

{\it Conclusions.}
The quantum jump description for Markovian systems (MCWF) is widely accepted due to its straightforward nature and the simple physical picture that it provides. For non-Markovian systems, the NMQJ method
maps memory into reverse jumps that restore quantum superpositions.
Furthermore,  
our approach becomes equivalent to the standard MCWF method in the Markovian limit. 
In a broader view, the continuously growing interest in quantum information~\cite{Stenholm2005} and nanophysics~\cite{Wolf2006} emphasises the need to consider single quantum systems at diminishing time scales and in tailored and finite environments. This development provides the background for the NMQJ approach. 

\acknowledgments
This work has been supported by the Academy of Finland
(Projects No.~108699, No.~115682, and No.~115982)
and the Magnus Ehrnrooth Foundation. We thank H.-P. Breuer and B. Garraway
for stimulating discussions.

\end{document}